\documentclass{jpaper}

\usepackage[nocompress]{cite}
\usepackage{algorithmic}
\usepackage{array}

\makeatletter
\let\MYcaption\@makecaption
\makeatother

\usepackage[font=footnotesize]{subcaption}

\makeatletter
\let\@makecaption\MYcaption
\makeatother

\usepackage{fixltx2e}
\usepackage{dblfloatfix}
\usepackage[nolessnomore, italic]{mathastext}
\usepackage[T1]{fontenc}
\usepackage[usenames,dvipsnames,svgnames,table]{xcolor}
\usepackage[normalem]{ulem}
\usepackage{enumitem}
\usepackage{setspace}
\usepackage{indentfirst}
\usepackage{footmisc}
\usepackage{fancyhdr}
\usepackage{authblk}
\usepackage{url}
\usepackage[binary-units=true]{siunitx}
\usepackage[us,12hr]{datetime}
\usepackage[keeplastbox]{flushend}
\usepackage[hidelinks]{hyperref}

\usepackage{xspace}
\usepackage{algorithm2e}
\usepackage{graphicx}
\usepackage{booktabs}
\usepackage{multirow}

\newif\ifcameraready
\camerareadytrue

\newcommand{\versionnum}[0]{4}

\ifcameraready
\else
\fi

\ifcameraready
  \newcommand{\todo}[1][]{}
  \newcommand{\ch}[0]{}
\else
  \newcommand{\todo}[1][]{\textbf{\fcolorbox{black}{red}{\color{white}{TODO}}} \underline{$\overline{\hbox{\emph{#1}}}$}}
  \newcommand{\ch}[1]{{\color{BrickRed} #1}}
\fi

\newcommand{\arsr}{\emph{alone-request-service-rate}\xspace}
\newcommand{\srsr}{\emph{shared-request-service-rate}\xspace}
\newcommand{\arsrabb}{\texttt{ARSR}\xspace}
\newcommand{\srsrabb}{\texttt{SRSR}\xspace}
\newcommand{\miiseabb}{MISE\xspace}

\newcommand{\miisefair}{MISE-Fair\xspace}
\newcommand{\miiseqos}{MISE-QoS\xspace}
\newcommand{\aoi}{AoI\xspace}

\newcommand{\ap}{\emph{Always Prioritize}\xspace}
\newcommand{\red}[0]{}

\title{Predictable Performance and Fairness Through Accurate\\
Slowdown Estimation in Shared Main Memory Systems}

\author{%
{Lavanya Subramanian$^{1,2}$}%
\qquad%
{Vivek Seshadri$^{3,2}$}%
\vspace{2pt}\\%
{Yoongu Kim$^{2}$}%
\qquad%
{Ben Jaiyen$^{4,2}$}%
\qquad%
{Onur Mutlu$^{5,2}$}%
}
\affil{{\em%
$^1$Intel Labs%
\qquad%
$^2$Carnegie Mellon University%
\qquad%
$^3$Microsoft Research India%
\qquad%
$^4$Google%
\qquad%
$^5$ETH Z{\"u}rich}%
}

\begin{document}
\maketitle



\begin{abstract}
\red{%
This paper summarizes the ideas and key concepts of MISE (Memory
Interference-induced Slowdown Estimation), which was published in
HPCA 2013~\cite{mise}, and examines the work's significance and
future potential. Applications running concurrently on
a multicore system interfere with each other at the main memory.
This interference can slow down different applications
differently. Accurately estimating the slowdown of each
application in such a system can enable mechanisms that can
enforce quality-of-service. While much prior work has focused on
mitigating the performance degradation due to inter-application
interference, there is little work on \ch{accurately} estimating slowdown of
individual applications in a multi-programmed environment. 
Our goal is to accurately estimate
application slowdowns, towards providing predictable performance.}

\red{%
To this end, we first build a simple Memory Interference-induced
Slowdown Estimation (MISE) model, which accurately estimates
slowdowns caused by memory interference.  We then leverage our
MISE model to develop two new memory scheduling schemes: 1) one
that provides soft quality-of-service guarantees, and 2) another
that explicitly attempts to minimize maximum slowdown (i.e.,
unfairness) in the system. Evaluations show that our techniques
perform significantly better than state-of-the-art memory
scheduling approaches to address the above problems.}

\ch{Our proposed model and techniques have enabled significant research
in the development of accurate performance models~\cite{asm, sem, jahre.hpca18,
li.cluster17}
and interference management mechanisms~\cite{mitts, camouflage,
afp, bliss-iccd14, bliss-tpds, afp, xiang.ics17}.}
\end{abstract}

\section{Problem: Unpredictable Slowdowns}
In a multicore system, multiple applications are consolidated on
the same machine. While consolidation may enable better resource
utilization, it results in interference between applications at
the shared resources, slowing down each application to a different
degree. Specifically, main memory is a heavily contended shared
resource between applications in a multicore system. Each
application accessing the memory experiences different and
unpredictable slowdowns depending on the available memory
bandwidth and the other concurrently running applications.

A large body of work proposed several different approaches to
mitigate memory interference between applications with the goal of
improving overall system performance. This includes memory
scheduling\ch{~\cite{sms,atlas,tcm,stfm,parbs,rlmc,fqm,bliss-iccd14,bliss-tpds,dash-taco,mph,saugata-isca13,pams,firm,lee.micro08}},
memory channel/bank partitioning\red{~\cite{bank-part,
mcp,bank-part-pact12}}, memory interleaving~\cite{mop}, source
throttling\ch{~\cite{fst,thottethodi-hpca01,baydal-tpds,hat, eiman-isca11}},
\ch{and} thread
scheduling\ch{~\cite{tang-isca11,zhuravlev-asplos10,adrm, a2c}} techniques.
However, few previous works
(notably~\cite{cycle-accounting-taco,fst,stfm,eiman-isca11}) have
attempted to estimate individual application slowdowns online with
the goal of providing predictable performance.

\textbf{Our goal} in \red{our HPCA 2013 paper~\cite{mise}} is to
provide predictable performance for individual applications. To
this end, we first design a model to accurately estimate
memory-interference-induced slowdowns of applications running
concurrently on a multicore system. We then leverage this model to
design effective mechanisms to enforce quality-of-service (QoS)
and achieve fairness.

\section{The Memory Interference-Induced\\Slowdown Estimation
(MISE) Model}
The slowdown of an application \red{indicates} the performance of the
application, when it is sharing resources with other applications,
relative to when the application is run alone. Slowdown can be
expressed as 
\begin{equation}
  \textrm{Slowdown of an App.} =
  \frac{\textrm{\emph{alone-performance}}}{\textrm{\emph{shared-performance}}}
  \label{eqn:basic-slowdown}
\end{equation}
Hence, estimating the slowdown of an individual application
requires two pieces of information: 1)~the performance of the
application when it is run concurrently with other applications
\red{(i.e., \emph{shared-performance})}, and 2)~the performance of
the application when it is run alone on the same system
\red{(i.e., \emph{alone-performance})}. While the former can be
directly measured, the key challenge is to estimate the
performance the application would have if it were running alone
\emph{while} it is actually running alongside other applications.
This requires quantifying the effect of interference on
application performance.

\subsection{Key Observations}

In this work, we make two observations
that lead to a simple and effective model to estimate the slowdown
of individual applications.

\textbf{Observation 1}: \emph{The performance of a memory-bound
application is roughly proportional to the rate at which its
memory requests are served.} This observation stems from
a memory-bound application's characteristic to spend an
overwhelmingly large fraction of its execution time stalling on
memory accesses. Therefore, the rate at which such an
application's requests are served has significant impact on its
performance. 

\begin{figure}[h]
    \centering 
    \includegraphics[scale=0.29, angle=270]{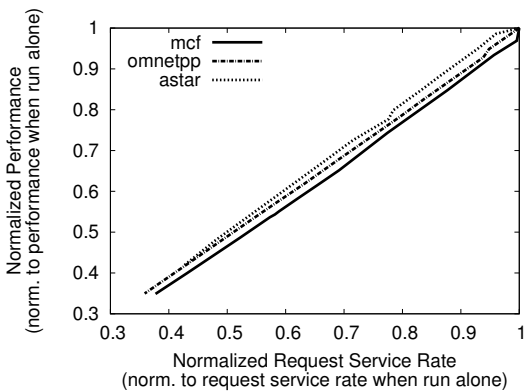}
    \caption{{Request service rate vs.\ performance. \red{Reproduced from~\cite{mise}.}}}
    \label{fig:bwfraction-speedup}
\end{figure}

To validate this observation, we conducted a real-system
experiment where we ran a memory-bound application from the SPEC
CPU2006 benchmark suite~\cite{spec2006} alongside three copies of
a microbenchmark whose memory intensity can be varied, on a 4-core
Intel Core i7~\cite{nehalem}.\footnote{The microbenchmark streams
through a large region of memory (one block at a time). The memory
intensity of the microbenchmark (\red{last-level cache
misses per kilo-instruction, i.e., LLC MPKI}) is varied by
changing the amount of computation performed between memory
operations.} By varying the memory intensity, \red{i.e., the
last-level cache (LLC) miss rate,} of the microbenchmark, we can
change the rate at which the requests of the SPEC application are
served.  Figure~\ref{fig:bwfraction-speedup} plots the results of
this experiment for three memory-intensive benchmarks, \emph{mcf},
\emph{omnetpp}, and \emph{astar}. The figure shows the performance
of each application versus the rate at which its requests are
served. 
 
The results of this experiment validate our observation. The
performance of a memory-bound application is directly proportional
to the rate at which its requests are served. This suggests that
we can use the request-service-rate of an application as a proxy
for its performance. More specifically, we can estimate the
slowdown of an application, i.e., the ratio of its performance
when it is run alone on a system vs. its performance when it is
run alongside other applications on the same system, as follows:
\begin{equation}
  \textrm{Slowdown of an App.} =
  \frac{\textrm{\arsr}}{\textrm{\srsr}}
  \label{eqn:slowdown-memory-bound}
\end{equation}
Estimating the \srsr (\srsrabb) of an application is
straightforward. It only requires the memory controller to keep
track of how many requests of the application are served in
a given number of cycles. However, the challenge is to estimate
the \arsr (\arsrabb) of an application \emph{while} it is run
alongside other applications. A naive way of estimating \arsrabb
of an application would be to prevent all other applications from
accessing memory for a length of time and measure the
application's \arsrabb. While this approach might provide an
estimate of the application's \arsrabb, it would significantly
slow down other applications in the system and is prone to
\red{incorrect estimations due to} phase fluctuations \red{in the
application}. Our second observation helps us to address this
problem.

\textbf{Observation 2}: \emph{The \arsrabb of an application can
be estimated by giving the requests of the application the
highest priority in accessing memory.}

Giving an application's requests the \emph{highest priority} in
accessing memory results in very little interference from the
requests of other applications. Therefore, requests of the
application are served almost as if the application were the only
one running on the system. Based on the above observation, the
\arsrabb of an application can be estimated as:
\begin{equation}
  \textrm{\arsrabb of an App.} = \frac{\textrm{\# Requests with
      Highest Priority}}{\textrm{\# Cycles with Highest Priority}}
  \label{eqn:arsr}
\end{equation}
where \emph{\# Requests with Highest Priority} is the number of
requests served when the application is given highest priority,
and \emph{\# Cycles with Highest Priority} is the number of cycles
an application is given highest priority by the memory controller. 

The memory controller can use Equation~\ref{eqn:arsr} to
periodically estimate the \arsrabb of an application. We add an
interference counter to capture the remaining interference cycles.
\red{The details of the mechanisms we add to increase the accuracy
of the model are described in Section 4 of our HPCA 2013 paper~\cite{mise}.}
Once we estimate \arsrabb,
Equation~\ref{eqn:slowdown-memory-bound} can be used to \red{estimate}
the slowdown of the application.

\subsection{MISE Model for Non-Memory-Bound\\Applications} So far,
we have described the key observations of the \miiseabb model for
a memory-bound application. We find that the model presented above
has low accuracy for non-memory-bound applications. This is
because a non-memory-bound application spends a significant
fraction of its execution time in the {\em compute phase} (when
the core is \red{\emph{not}} stalled waiting for memory). Hence,
varying the request service rate for such an application will not
affect the length of the large compute phase.  Therefore, we take
into account the duration of the compute phase to make the model
accurate for non-memory-bound applications.

Let $\alpha$ be the fraction of time spent by an application \red{stalling at
memory.}
Therefore, the fraction of time spent by the
application in the compute phase is $1-\alpha$. Since changing the
request service rate affects only the memory phase, we augment
Equation~\ref{eqn:slowdown-memory-bound} to take into account
$\alpha$ as follows:
\begin{equation}
  \emph{Slowdown of an App.} = (1 - \alpha) + \alpha
  \frac{\arsrabb}{\srsrabb}
  \label{eqn:slowdown-non-memory-bound}
\end{equation}

In addition to estimating \arsrabb and \srsrabb required by
Equation~\ref{eqn:slowdown-memory-bound}, the above equation
requires estimating the parameter $\alpha$, the fraction of time
spent in \red{the} memory phase. However, precisely computing $\alpha$
for a modern out-of-order processor is a challenge since such a
processor overlaps computation with memory accesses. The processor
stalls waiting for memory only when the oldest instruction in the
reorder buffer is waiting on a memory request. For this reason,
we estimate $\alpha$ as the fraction of time the processor spends
stalling for memory:
\begin{equation}
  \alpha = \frac{\textrm{\# Cycles spent stalling on memory
      requests}}{\textrm{Total number of cycles}}
  \label{eqn:alpha}
\end{equation}

\red{More details of our MISE slowdown estimation model are
described in Sections~3 and 4 of our HPCA 2013 paper~\cite{mise}. More recently, we used
this model to expand slowdown estimation to a memory hierarchy
that also includes shared caches, as part of the Application
Slowdown Model~\cite{asm}.} 

\section{Evaluation of the \miiseabb Model}
\label{sec:model-evaluation}

We compare the \miiseabb model against the slowdown estimation
model employed by the Stall Time Fair Memory Scheduler
(STFM)~\cite{stfm}, which is the closest previous work on
estimating memory interference-induced
slowdown.\footnote{FST~\cite{fst} and Du Bois et al.'s \red{per-thread cycle accounting}
mechanism~\cite{cycle-accounting-taco} are the other two previous
works that estimate application slowdown. The mechanism to
estimate main memory interference induced slowdown in both of
these previous works is similar to STFM.} STFM estimates the
slowdown of an application by estimating the number of cycles it
stalls due to interference from other applications' requests. In
this section, we qualitatively and quantitatively compare
\miiseabb with STFM.
 
There are two key differences between \miiseabb and STFM in
estimating slowdown. First, \miiseabb uses request service rates
rather than stall times to estimate slowdown. \red{In MISE,} the
\arsr of an application can be fairly accurately estimated by
giving the application highest priority in accessing memory.
Giving the application highest priority in accessing memory
results in very little interference from other applications. In
contrast, STFM attempts to estimate the alone-stall-time of an
application while it is receiving significant interference from
other applications, \red{which turns out to be difficult to do
accurately.} Second, \miiseabb takes into account the effect of
the compute phase for non-memory-bound applications. STFM, on the
other hand, has no such provision to account for the compute
phase. As a result, \miiseabb's slowdown estimates for
non-memory-bound applications are significantly more accurate than
STFM's estimates.

\begin{figure}[!htb]
  \centering
  \begin{subfigure}{0.75\linewidth}
      \centering
      \includegraphics[width=\linewidth, angle=0]{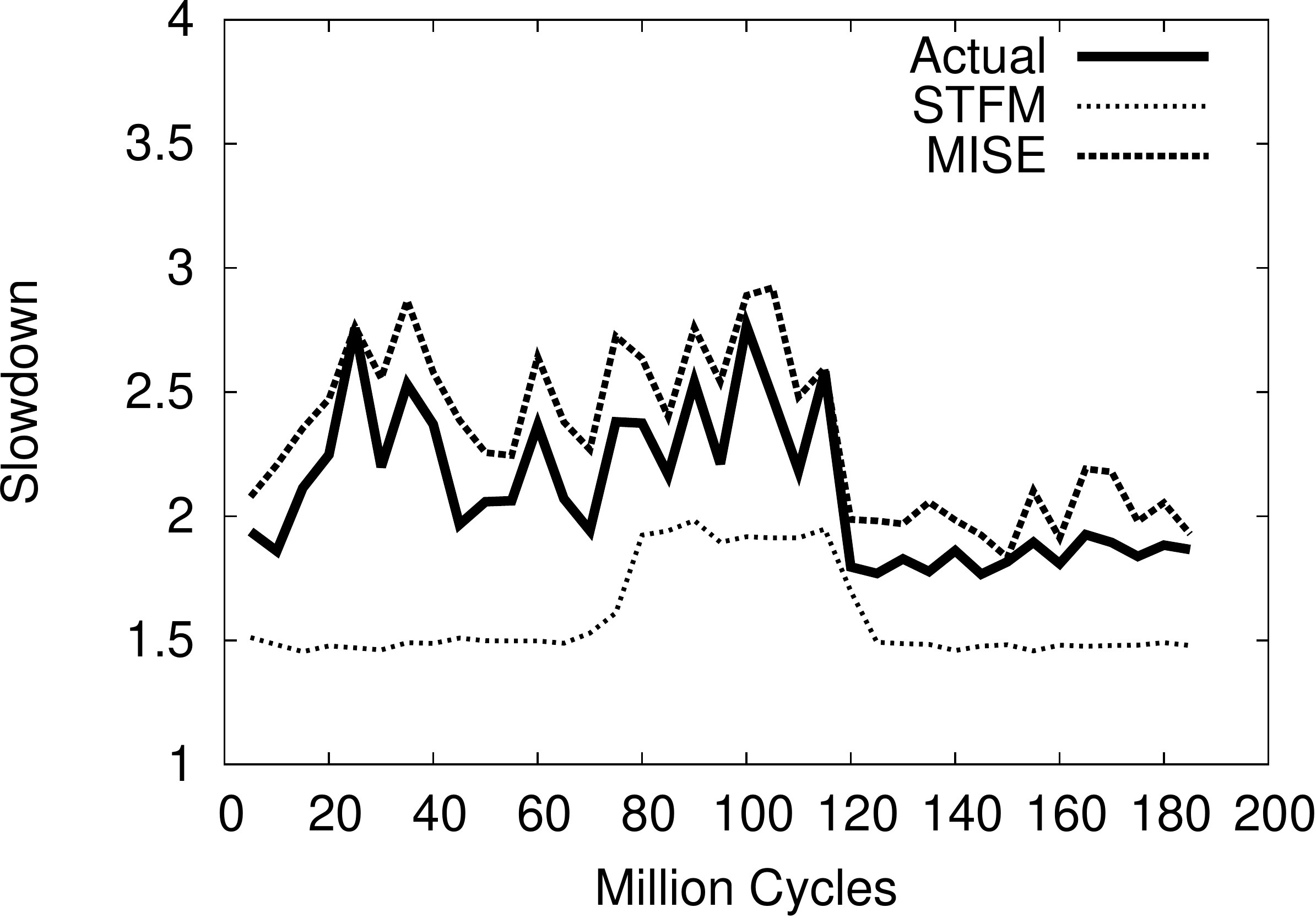}
      \subcaption{lbm}
      \label{fig:lbm-stfm-comp}
  \end{subfigure}
\\%
  \begin{subfigure}{0.75\linewidth}
      \centering
      \includegraphics[width=\linewidth, angle=0]{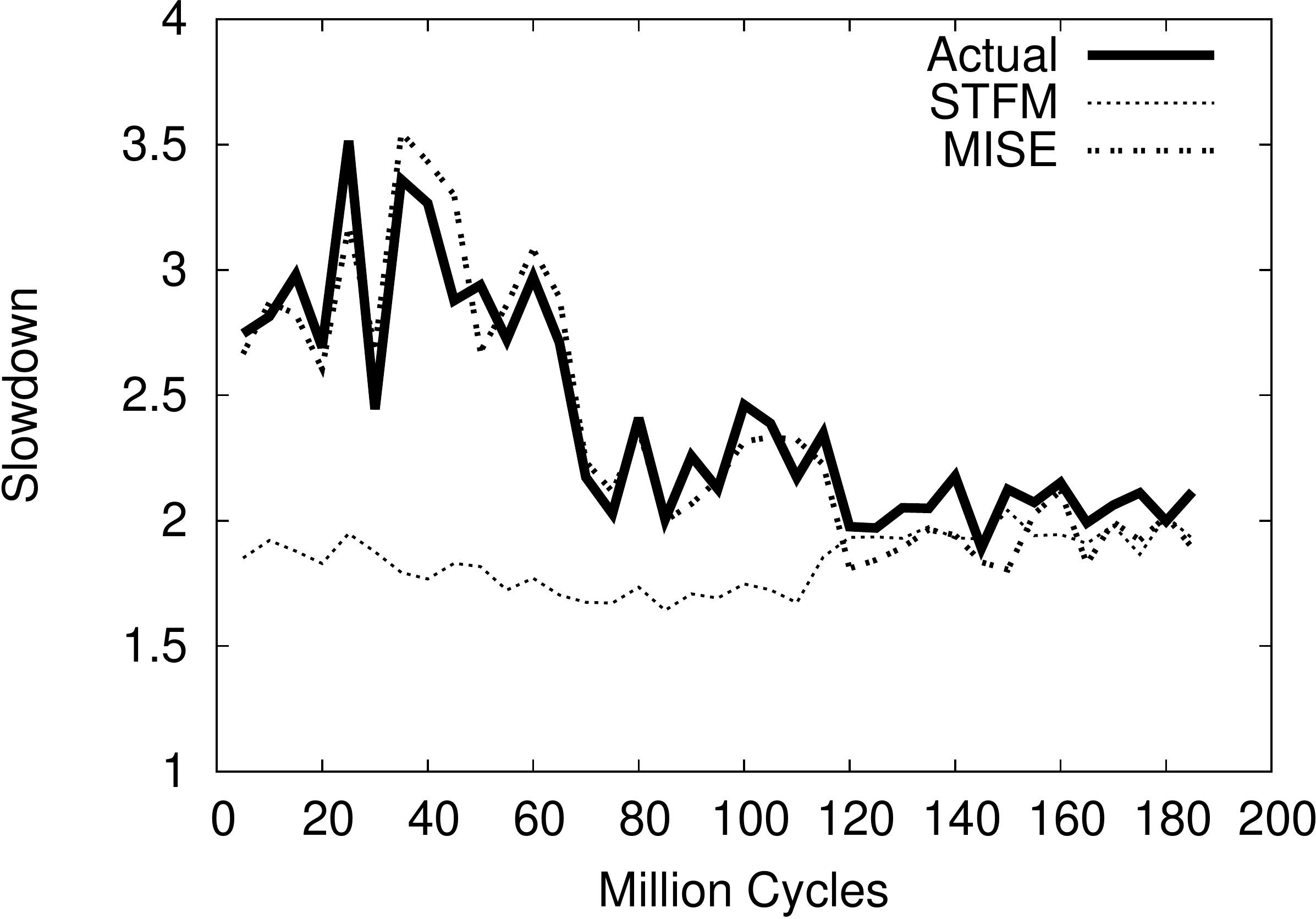}
      \subcaption{leslie3d}
      \label{fig:leslie-stfm-comp}
  \end{subfigure}
  \caption{Comparison of \miiseabb with STFM for representative memory-bound applications. \red{Adapted from~\cite{mise}.}}
  \label{fig:comparison-stfm-mb}
\end{figure}

\begin{figure}[!htb]
  \centering
  \begin{subfigure}{0.75\linewidth}
      \centering
      \includegraphics[width=\linewidth, angle=0]{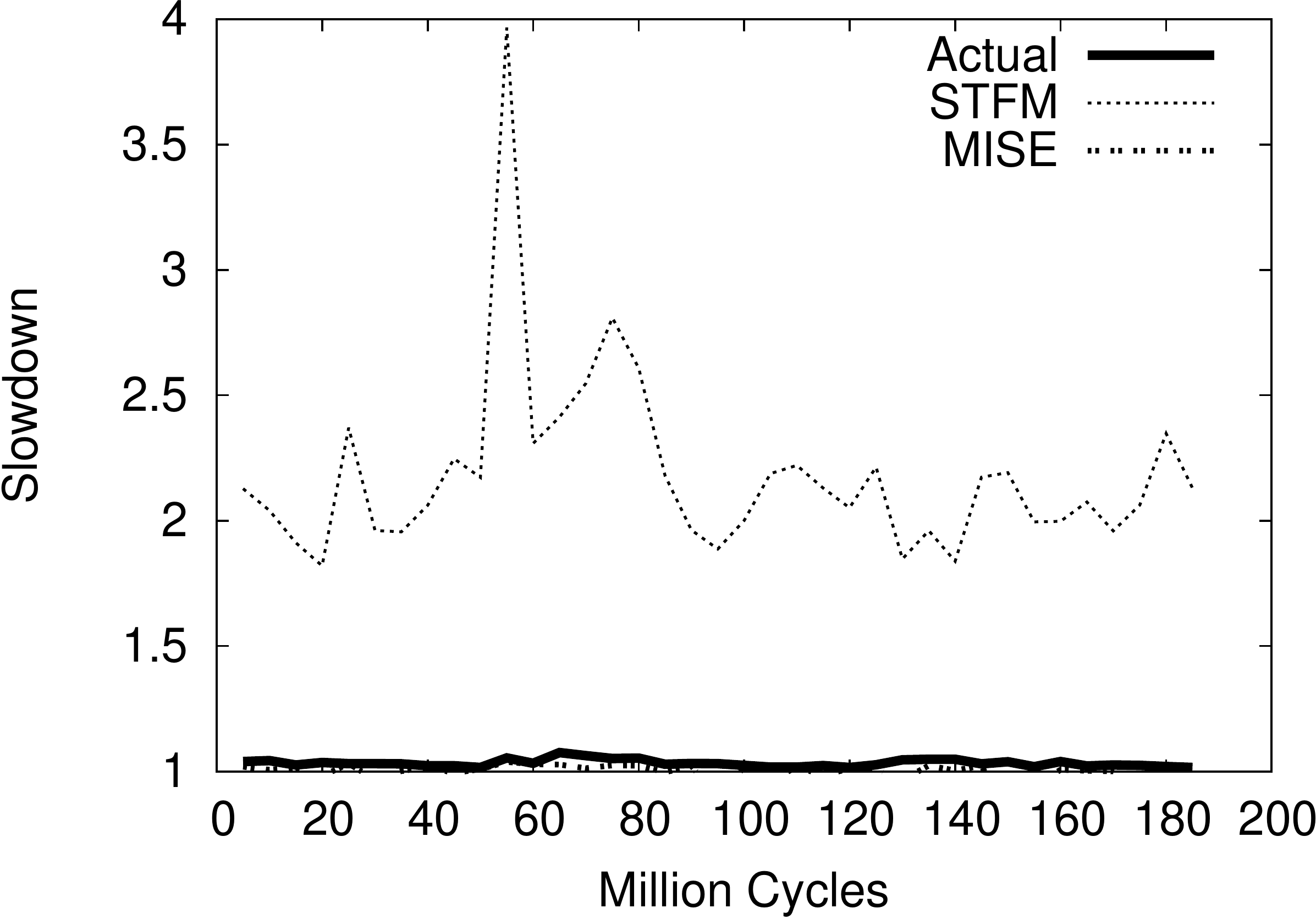}
      \subcaption{wrf}
      \label{fig:calculix-stfm-comp}
  \end{subfigure}%
  \\
  \begin{subfigure}{0.75\linewidth}
      \centering
      \includegraphics[width=\linewidth, angle=0]{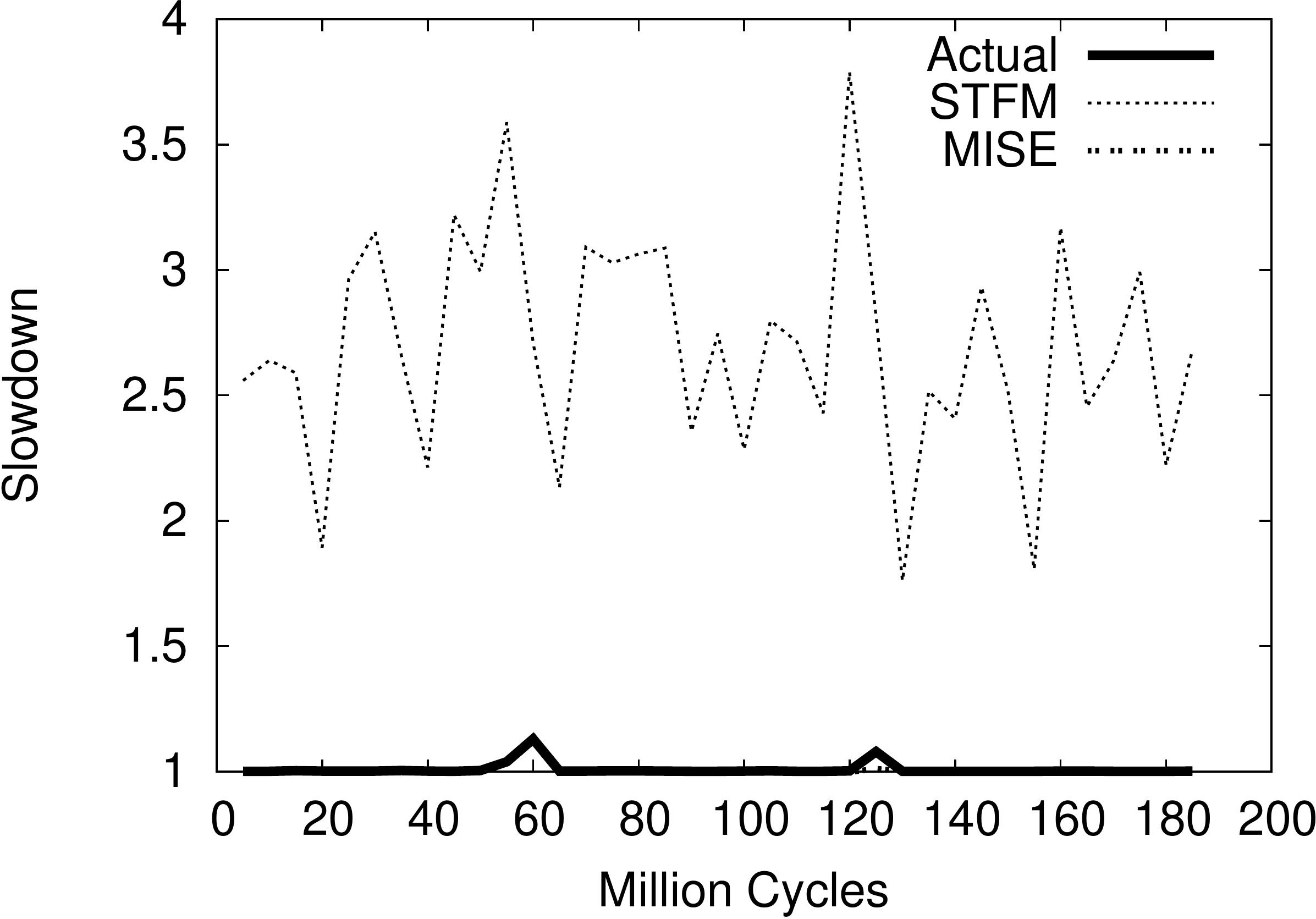}
      \subcaption{povray}
      \label{fig:povray-stfm-comp}
  \end{subfigure}
  \caption{Comparison of \miiseabb with STFM for representative non-memory-bound applications. \red{Adapted from~\cite{mise}.}}
  \label{fig:comparison-stfm-nonmb}
\end{figure}

Figure~\ref{fig:comparison-stfm-mb} compares the accuracy of
\miiseabb with STFM for two representative memory-bound
applications, lbm and leslie3d.
Figure~\ref{fig:comparison-stfm-nonmb} compares the accuracy of
\miiseabb with STFM for two representative non-memory-bound
applications, wrf and povray. Each of these applications is run on
a 4-core system with three other applications. \red{Our detailed
experimental methodology is \ch{provided} in Section 5 of our HPCA 2013 paper~\cite{mise}.
This includes detailed descriptions of our experimental setup,
workloads and metrics. Furthermore, our simulator implementing the
MISE model is available online~\cite{asmsim}.} As can be
observed, \miiseabb's slowdown estimates are much closer to the
actual slowdown than STFM's estimates. This is because the
\miiseabb model eliminates a significant portion of the
interference received by an application while estimating slowdown,
by prioritizing it \red{in the memory controller}. On the other
hand, STFM estimates slowdown \emph{while} an application is
experiencing interference.

Table~\ref{tab:slowdown-errors} shows the average slowdown
estimation error for each benchmark, with STFM and \miiseabb,
across 300 4-core workloads of different memory intensities. As
can be observed, \miiseabb's slowdown estimates have significantly
lower error than STFM's slowdown estimates across most benchmarks.
Across {\em 300} workloads, STFM's estimates deviate from the
actual slowdown by 29.8\%, whereas, our proposed \miiseabb model's
estimates deviate from the actual slowdown by only 8.1\%.
Therefore, we conclude that our slowdown estimation model provides
better accuracy than STFM.

\begin{table}[h]
  \centering
  \setlength{\tabcolsep}{0.45em}
  \vspace{5pt}
  \caption{Average \red{slowdown estimation error} for each benchmark (in \%). \red{Adapted from~\cite{mise}.}}
  \label{tab:slowdown-errors}
  \footnotesize{
\begin{tabular}{|c|c|c||c|c|c|}
  \hline
  \textbf{Benchmark} & \textbf{STFM} & \textbf{\miiseabb\xspace} & \textbf{Benchmark} & \textbf{STFM} & \textbf{\miiseabb\xspace}\\
  \hline
  \hline
  453.povray & 56.3 & 0.1 & 473.astar & 12.3 & 8.1\\
  \hline
  454.calculix & 43.5 & 1.3 & 456.hmmer & 17.9 & 8.1\\
  \hline
  400.perlbench	& 26.8 & 1.6 & 464.h264ref & 13.7 & 8.3\\
  \hline
  447.dealII & 37.5 & 2.4 & 401.bzip2 &	28.3 & 8.5\\
  \hline
  436.cactusADM	& 18.4 & 2.6 & 458.sjeng & 21.3 & 8.8\\
  \hline
  450.soplex & 29.8 & 3.5 & 433.milc & 26.4 & 9.5\\
  \hline
  444.namd & 43.6 & 3.7 & 481.wrf & 33.6 & 11.1\\
  \hline
  437.leslie3d & 26.4 & 4.3 & 429.mcf &	83.74 & 11.5\\
  \hline
  403.gcc & 25.4 & 4.5 & 445.gobmk & 23.1 & 12.5\\
  \hline
  462.libquantum & 48.9 & 5.3 & 483.xalancbmk &	18.0 & 13.6\\
  \hline
  459.GemsFDTD & 21.6 & 5.5 & 435.gromacs & 31.4 & 15.6\\
  \hline
  470.lbm & 6.9 & 6.3 & 482.sphinx3 & 21 & 16.8\\
  \hline
  473.astar & 12.3 & 8.1 & 471.omnetpp & 26.2 & 17.5\\
  \hline
  456.hmmer & 17.9 & 8.1 & 465.tonto & 32.7 & 19.5\\
  \hline
\end{tabular}
}

\end{table}

\red{For a more detailed analysis of the MISE model's accuracy and characteristics, we refer the reader to
our HPCA 2013 paper~\cite{mise}.} 

\section{Leveraging the \miiseabb Model}
\label{sec:leveraging}
Accurate slowdown estimates are a key enabler towards designing
mechanisms to better enforce quality-of-service (QoS) and
fairness. Slowdown estimates from the \miiseabb model could be
leveraged in hardware to design memory scheduling policies to
provide QoS guarantees and fairness. Alternatively, the slowdown
estimates could be communicated to the system software, which
could leverage them to perform application scheduling, admission
control and migration. We will describe two such mechanisms that
leverage the \miiseabb model: 1) \miiseqos, a mechanism to provide
soft QoS guarantees in the memory controller; and 2) \miisefair,
a mechanism to minimize maximum slowdown\red{~\cite{a2c,
dash-taco, app-aware-prio, atlas, tcm, bliss-iccd14, bliss-tpds,
eaf-cache}} to improve overall system fairness. 



\subsection{\miiseqos: Providing Soft QoS Guarantees}
\miiseqos aims to provide soft slowdown guarantees to an
application of interest (\aoi) in a workload with many applications,
while trying to maximize overall performance for the remaining
applications. There are two aspects of providing a soft slowdown
guarantee. One is to ensure that the application of interest is
not slowed down by more than a system-software-specified bound.
The other aspect is to detect if the bound is \emph{not met} for
some reason. 

\miiseqos addresses both of these aspects by using slowdown
estimates from the \miiseabb model. It periodically obtains
slowdown estimates from the \miiseabb model and
increases/decreases the memory bandwidth allocated to the \aoi
such that the \aoi receives just enough bandwidth to meet its
slowdown bound. This enables the other applications to use the
remaining bandwidth, improving their performance. \miiseqos
addresses the second aspect by comparing slowdown estimates from
the \miiseabb model with the prescribed bound periodically. When
the prescribed bound cannot be met despite always prioritizing the
\aoi, \miiseqos detects that the bound cannot be met just by
prioritizing the application at the memory controller.

Previous work~\cite{qos-sigmetrics} attempts to address the first
aspect by \red{\emph{always}} prioritizing the \aoi. This may
unnecessarily slowdown other applications in the system by
excessively prioritizing the \aoi, \red{especially when the \aoi
is meeting its performance \ch{bound}}. Furthermore, such a mechanism,
in the absence of accurate slowdown estimates, does not have the
provision to detect whether or not the bound is met.


\textbf{Slowdown Evaluation.}
We evaluate the \miiseqos mechanism across 300 workloads with 10
different slowdown bounds for each workload. Our results show that
the \miiseqos mechanism meets the prescribed slowdown bound for
97.5\% of the workloads for which the naive mechanism that always
prioritizes the \aoi meets the bound, while improving overall
system performance by 12\%.  Furthermore, \miiseqos also predicts
whether or not the bound is met with an accuracy of 95.7\%, while
previous work~\cite{qos-sigmetrics} has no such provision.

\begin{figure}[h]
  \centering
  \includegraphics[scale=0.35, angle=270]{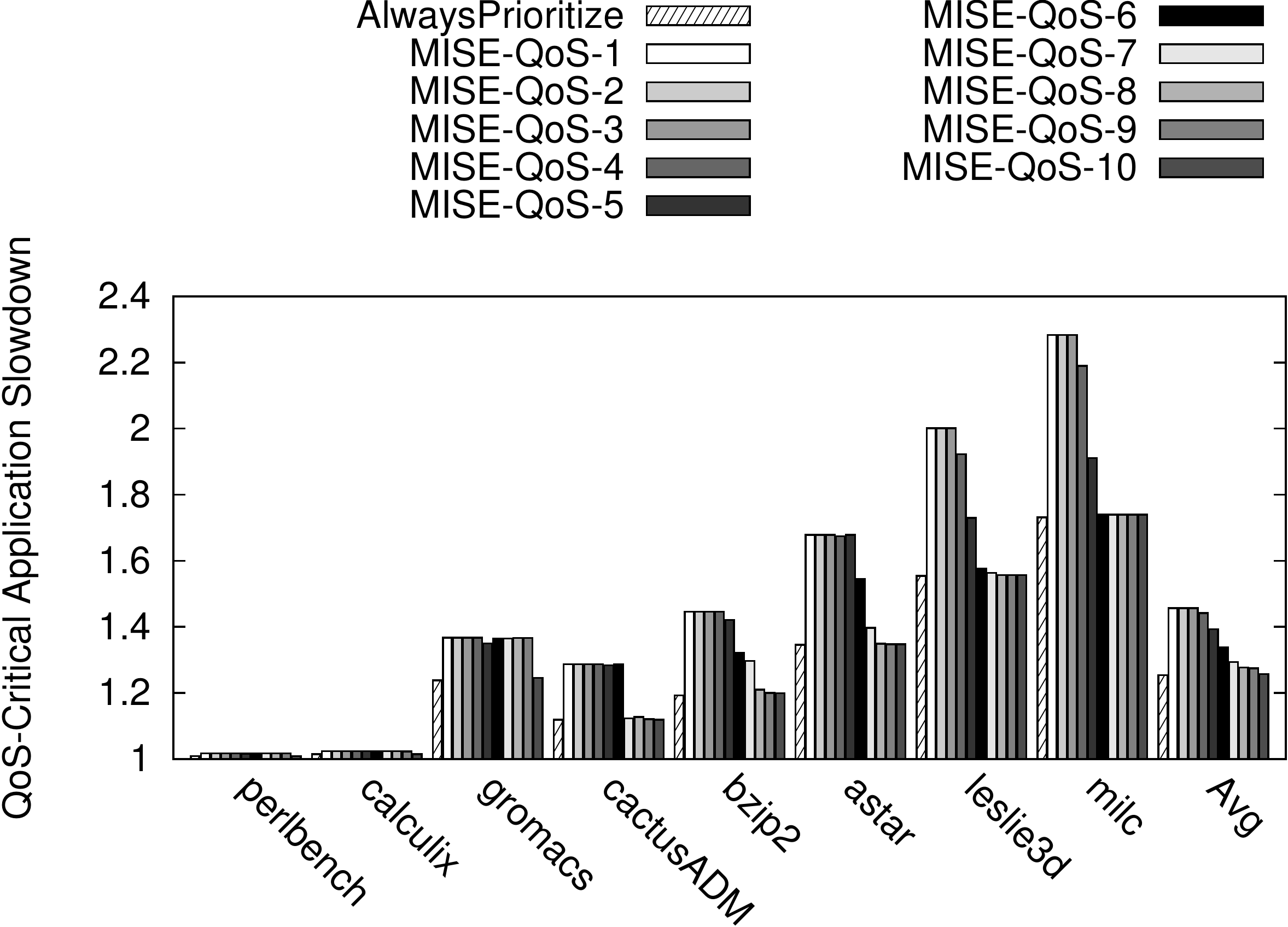}
  \caption{\aoi performance: \miiseqos vs. \emph{AlwaysPrioritize}. \red{Reproduced from~\cite{mise}.}}
  \label{fig:ind-app-slowdown}
\end{figure}

To show the effectiveness of \miiseqos, we compare the AoI's
slowdown due to \miiseqos and the mechanism that always
prioritizes the AoI (\ap)~\cite{qos-sigmetrics}.
Figure~\ref{fig:ind-app-slowdown} presents representative results
for 8 different AoIs when they are run alongside three other
applications. The label \miiseqos-n corresponds to a slowdown
bound of $\frac{10}{n}$. (Note that \ap does not take into account
the slowdown bound.) Note that the slowdown bound decreases (i.e.,
becomes tighter) from left to right for each benchmark in
Figure~\ref{fig:ind-app-slowdown} (as well as in other figures).

\begin{figure*}[!ht] 
  \centering 
  \begin{minipage}{0.45\textwidth}
    \centering 
    \includegraphics[scale=0.29, angle=270]{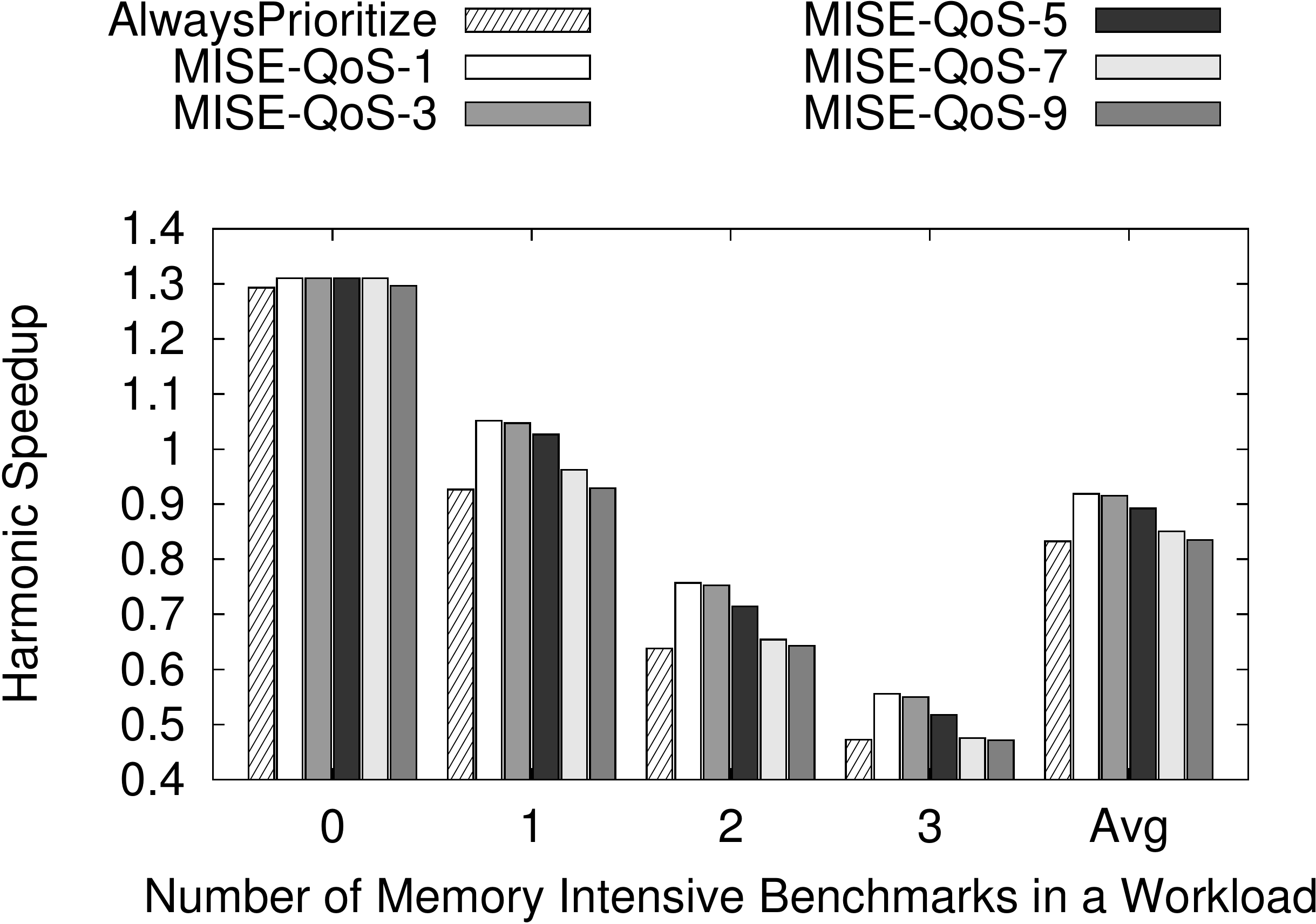}
  \end{minipage} 
  \begin{minipage}{0.45\textwidth}
    \centering
    \includegraphics[scale=0.29, angle=270]{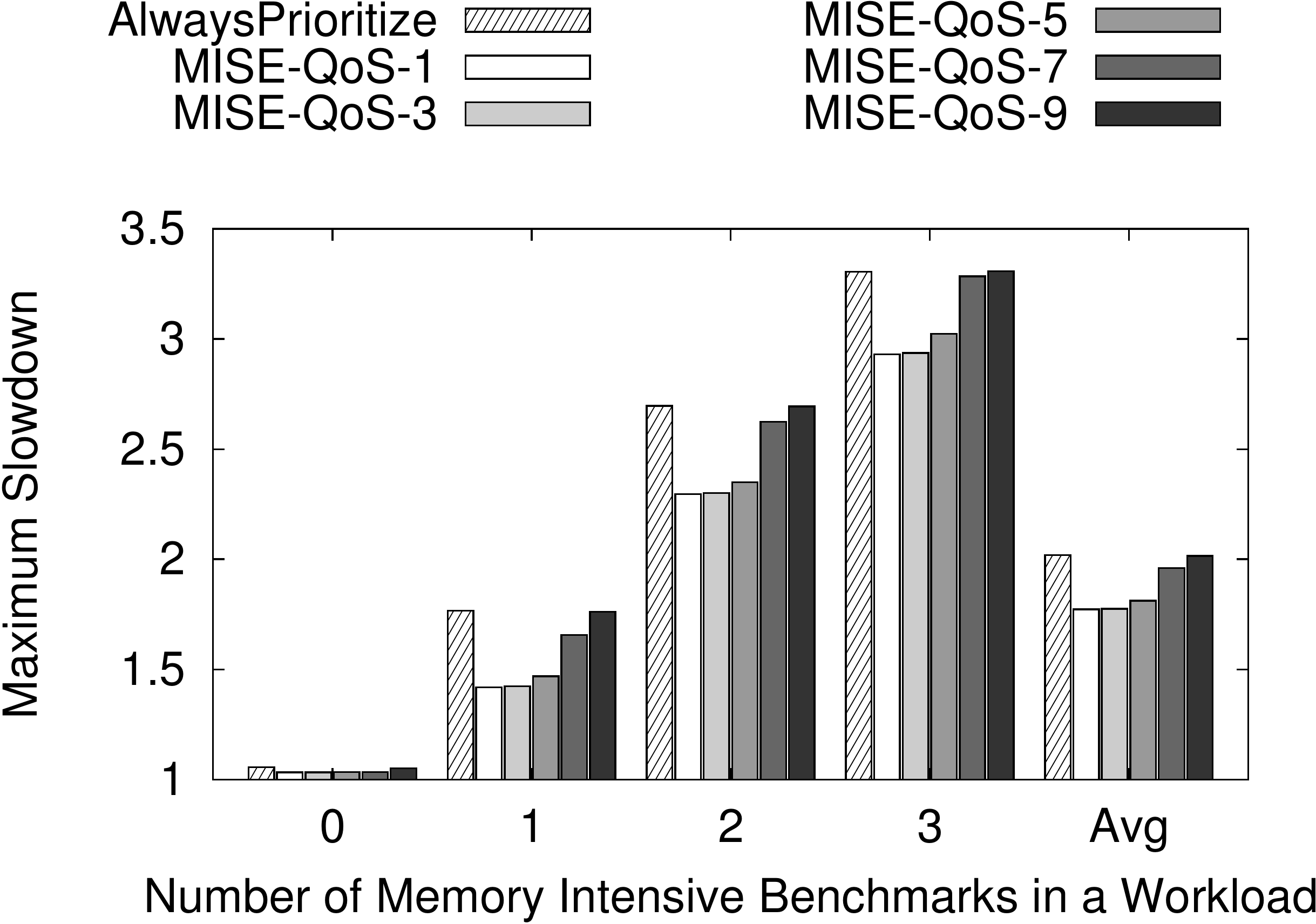}
  \end{minipage}
  \caption{Average system performance and fairness across
    300 workloads of different memory intensities. \red{Reproduced from~\cite{mise}.}}
  \label{fig:sef-avg-hs-ms}
\end{figure*}

We draw three conclusions from the results.  First, for most
applications, the slowdown of \ap is considerably more than one.
\red{This indicates that} always prioritizing the AoI does
\red{not} completely prevent other applications from interfering
with the AoI.
Second, as the slowdown bound for the AoI is decreased (left to
right), \miiseqos gradually increases the bandwidth allocation for
the AoI, eventually allocating all the available bandwidth to the
AoI. At this point, \miiseqos performs very similarly to the
\ap mechanism.
Third, in almost all cases (in this figure and across all our 3000
data points), \miiseqos meets the specified slowdown bound
\emph{if} \ap is able to meet the bound \red{(see Section~8.1 of our
HPCA 2013 paper~\cite{mise} for details)}.

\textbf{System Performance and Fairness.}
Figure~\ref{fig:sef-avg-hs-ms} compares the system performance
(harmonic speedup) and fairness (maximum slowdown) of \miiseqos
and \ap for different values of the bound. We omit the AoI from
the performance and fairness calculations. The results are
categorized into four workload categories (0, 1, 2, 3) indicating
the number of memory-intensive benchmarks in the workload. For
clarity, the figure shows results only for a few slowdown
bounds. Three conclusions are in order.

First, \miiseqos significantly improves performance compared to
\ap, especially when the slowdown bound for the AoI is large. On
average, when the bound is $\frac{10}{3}$, \miiseqos improves
harmonic speedup\red{~\cite{hmeanfair01}} by 12\% and weighted
speedup\red{~\cite{ws,sysperf}} by 10\% (not shown due to lack of
space) over \ap, while reducing maximum slowdown\red{~\cite{a2c,
dash-taco, app-aware-prio, atlas, tcm, bliss-iccd14, bliss-tpds,
eaf-cache}} by 13\%. Second, as expected, the performance and
fairness of \miiseqos approach that of \ap as the slowdown bound
is decreased (going from left to right for a set of bars).
Finally, the benefits of \miiseqos increase with increasing memory
intensity because always prioritizing a memory intensive
application will cause significant interference to other
applications.

Based on our results, we conclude that \miiseqos can effectively
ensure that the AoI meets the specified slowdown bound while
achieving high system performance and fairness across the other
applications. 

\subsection{\miiseabb-Fair: Minimizing Maximum Slowdown}
\label{sec:app2}

The second mechanism we build on top of our \miiseabb model is one
that seeks to improve overall system fairness. Specifically, this
mechanism attempts to minimize the maximum slowdown across all
applications in the system. Ensuring that no application is
unfairly slowed down while maintaining high system performance is
an important goal in multicore systems where co-executing
applications are similarly important. Many prior works evaluate
fairness in such scenarios in terms of the maximum slowdown of any
application\red{~\cite{a2c, dash-taco, app-aware-prio, atlas, tcm,
bliss-iccd14, bliss-tpds, eaf-cache}}.  

At a high level, our mechanism works as follows. The memory
controller maintains two pieces of information: 1)~a target
slowdown bound ($B$) for \red{\emph{all}} applications, and 2)~a
bandwidth allocation policy that partitions the available memory
bandwidth across all applications.  The memory controller enforces
the bandwidth allocation policy using a lottery-scheduling
technique proposed in~\cite{lottery-scheduling-waldspurger}.  The
controller attempts to ensure that the slowdown of all
applications is within the bound $B$. To this end, it modifies the
bandwidth allocation policy so that applications that are slowed
down more get more memory bandwidth. Should the memory controller
find that bound $B$ is not possible to meet, it increases the
bound. On the other hand, if the bound is easily met, it decreases
the bound. 


\textbf{Interaction with the Operating System.} 
As we will
show in Section~\ref{sec:app2-eval}, our mechanism provides the
best fairness compared to three state-of-the-art approaches for
memory request scheduling~\cite{atlas,tcm,stfm}. In addition to
this, there is another benefit to using our approach. Our
mechanism, based on the \miiseabb model, can accurately estimate
the slowdown of each application. Therefore, the memory controller
can potentially communicate the estimated slowdown information to
the operating system (OS). The OS can use this information to make
more informed scheduling and mapping decisions \red{in order} to
further improve system performance or fairness. Since prior memory
scheduling approaches do \red{\emph{not}} explicitly attempt to
minimize maximum slowdown by accurately estimating the slowdown of
individual applications, such a mechanism to interact with the OS
is \red{\emph{not}} possible with them. Evaluating the benefits of
the interaction between our mechanism and the OS is beyond the
scope of this paper \red{but is an important area of future work}.

\begin{figure}[!ht]
  \centering
  \includegraphics[scale=0.29, angle=270]{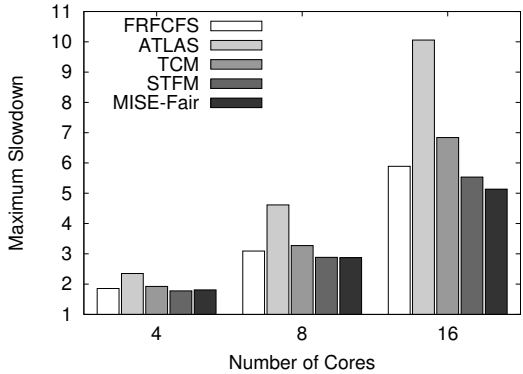}
  \caption{Fairness with different core counts. \red{Reproduced from~\cite{mise}.}}
  \label{fig:sef-app2-core-scalability-ms}
\end{figure}

\textbf{Evaluation.}
\label{sec:app2-eval}
Figure~\ref{fig:sef-app2-core-scalability-ms} compares the system
fairness (maximum slowdown) of different mechanisms with increasing
number of cores. The figure shows results with four previously
proposed memory scheduling policies (FRFCFS~\cite{frfcfs,
  frfcfs-patent}, ATLAS~\cite{atlas}, TCM~\cite{tcm}, and
STFM~\cite{stfm}), and our proposed mechanism using the \miiseabb
model (\miiseabb-Fair). We draw three conclusions from our results.

First, \miiseabb-Fair provides the best fairness compared to all
other previous approaches. The reduction in the maximum slowdown
due to \miiseabb-Fair when compared to STFM (the best previous
mechanism) increases with increasing number of cores. With 16
cores, \miiseabb-Fair provides 7.2\% better fairness compared to
STFM.

Second, STFM, as a result of prioritizing the most slowed down
application, provides better fairness than all other previous
approaches. While the slowdown estimates of STFM are not as
accurate as those of our mechanism, they are good enough to
identify the most slowed down application. However, as the number
of concurrently-running applications increases, simply
prioritizing the most slowed down application may not lead to
better fairness. \miiseabb-Fair, on the other hand, works towards
reducing maximum slowdown by stealing bandwidth from those
applications that are less slowed down compared to others. As a
result, the fairness benefits of \miiseabb-Fair compared to STFM
increase with increasing number of cores.

Third, ATLAS and TCM are more unfair compared to FRFCFS. As shown
in prior work~\cite{atlas,tcm}, ATLAS trades off fairness to
obtain better performance. TCM, on the other hand, is designed to
provide high system performance and fairness. Further analysis
showed us that the cause of TCM's unfairness is the strict ranking
employed by TCM. TCM ranks all applications based on its
clustering and shuffling techniques~\cite{tcm} and strictly
enforces these rankings. We found that such strict ranking
destroys the row-buffer locality of low-ranked applications. This
increases the slowdown of such applications, leading to high
maximum slowdown.\red{\footnote{\red{Note that this observation later
led us to develop the Blacklisting Memory Scheduler
(BLISS)~\cite{bliss-iccd14,bliss-tpds}.}}}

\red{We conclude that the MISE model's slowdown estimates can be
used to design a better and more fair memory scheduler. We expect
future works can take advantage of the MISE model to design even
better memory scheduling and other resource management
mechanisms.}


\section{Related Work}

To our knowledge, this is the first paper to 1) provide a {\em
simple and accurate} model to estimate application slowdowns in
the presence of main memory interference, and 2) use this model to
devise two new memory scheduling techniques that either aim to
satisfy slowdown bounds of applications or improve system fairness
and performance. 
In this section, we discuss several related works.
We discuss works that build upon MISE in Section~\ref{sec:retrospective}.

\textbf{Slowdown Estimation.} 
Stall Time Fair Memory Scheduling (STFM)~\cite{stfm}
attempts to estimate each application's slowdown, 
with the goal of improving fairness by prioritizing the
most slowed down application.  STFM estimates an application's 
slowdown as the ratio of its memory stall time when it
is run alone versus when it is concurrently run alongside other
applications. The challenge is in determining the 
alone stall time of an application \emph{while}
the application is actually running
alongside other applications.  STFM proposes to address this
challenge by counting the number of cycles an application is
stalled due to interference from other applications at the DRAM
channels, banks and row-buffers. STFM uses this interference
cycle count to estimate the alone-stall-time of the application,
and hence the application’s slowdown.

Fairness via Source Throttling (FST)~\cite{fst} estimates application
slowdowns due to inter-application interference at the shared
caches and memory, as the ratio of uninterfered to interfered
execution times. FST uses the slowdown estimates to make informed 
source throttling decisions, to improve fairness.  The
mechanism to account for memory interference to estimate uninterfered 
execution time is similar to that employed in STFM.
\red{Prefetch-Aware
Shared Resource Management~\cite{eiman-isca11} extends the FST
model to take into account prefetch requests.}

A concurrent work by Du Bois et al.~\cite{cycle-accounting-taco} 
proposes per-thread cycle accounting (PTCA) for multicore processors, 
which determines an application's standalone execution time
when it shares cache and memory with other applications in a
multicore system. In order to quantify memory interference, 
PTCA counts the number of waiting cycles due to inter-application 
interference and factors out
these waiting cycles to estimate alone execution times, which is
similar to STFM's alone stall time estimation mechanism.

Eyerman and Eeckhout~\cite{eeckhout-asplos2009} and Cazorla et
al.~\cite{pred-perf-smt} propose mechanisms to determine an
application's slowdown while it is running alongside other
applications on an SMT processor. Luque et
al.~\cite{pred-perf-cache} estimate application slowdowns in the
presence of shared cache interference. 
 Lin and
Balasubramonian~\cite{refining-utility-metric} propose a
regression-based model to estimate performance for different cache
allocations.
None of these studies take
into account inter-application interference at the main memory. Therefore,
\miiseabb, which estimates slowdown due to main memory
interference, can be combined with the above approaches
to quantify interference at the SMT processor and shared cache to
build a comprehensive mechanism.

\textbf{Quality-of-Service (QoS).}  Several prior works 
provide QoS guarantees in shared memory CMP systems. Mars et
al.~\cite{bubbleup} propose a mechanism to estimate an
application's sensitivity towards interference and its propensity
to cause interference. They utilize this knowledge to make
informed mapping decisions between applications and
cores. However, this mechanism 1)~assumes {\em a priori} knowledge
of applications, which may not always be possible to have, and
2)~is designed for only 2 cores, and it is not clear how it can be
extended to more than 2 cores. In contrast, \miiseabb does not
assume any {\em a priori} knowledge of applications and works well
with large core counts, as we have shown in this paper. That
said, \miiseabb can possibly be used to provide feedback to the
mapping mechanism proposed by~\cite{bubbleup} to overcome the
shortcomings of their mechanism.

Iyer et
al.~\cite{ratebasedqos,cqos,qos-sigmetrics} propose mechanisms to
provide guarantees on shared cache space, memory bandwidth or IPC
for different applications. The slowdown guarantee provided by
\miiseqos is stricter than these mechanisms as \miiseqos takes into
account the alone-performance of each application.  Nesbit et
al.~\cite{fqm} propose a mechanism to enforce a bandwidth
allocation policy, by partitioning the available bandwidth across
concurrently running applications based on some policy.  While we
use a scheduling technique similar to
lottery-scheduling~\cite{lottery-scheduling,
  lottery-scheduling-waldspurger} to enforce the bandwidth
allocation policies of \miiseqos and \miisefair, the mechanism
proposed by Nesbit et al.\ can also be used in our proposal to
allocate bandwidth instead of our lottery-scheduling approach.


\textbf{Memory Interference Mitigation.}  Many prior
works focus on the problem of mitigating inter-application
interference at the main memory to improve system performance and/or
fairness. Most of \ch{these} approaches address \ch{memory interference} by
modifying the memory request scheduling
algorithm\ch{~\cite{mph,sms,pams,rlmc,atlas,tcm,stfm,parbs,fqm, saugata-isca13,qos-sigmetrics,podc,
firm, complexity-effective, bliss-iccd14,bliss-tpds ,lee.micro08, llcwb, lee.micro09, lee.tc11}}. We
quantitatively compare \miisefair to STFM~\cite{stfm},
ATLAS~\cite{atlas}, and TCM~\cite{tcm} in Section~\ref{sec:app2}, and show that \miisefair
provides better fairness than these prior approaches. 

Other works examine
approaches such as sub-row interleaving~\cite{mop}, channel/bank
partitioning~\cite{mcp,bank-part,bank-part-pact12,bank-part-hpca2014},
bandwidth partitioning~\cite{mise,solihin-hpca2010}, source
throttling\ch{~\cite{fst,hat,nychis,cc-hotnets10,thottethodi-hpca01,
baydal-tpds,kayiran-micro14, eiman-isca11}}, thread
scheduling~\cite{zhuravlev-asplos10,tang-isca11,a2c,adrm},
\ch{and changes to DRAM design~\cite{lee.pact15, salp}}.
These approaches are complementary to MISE, and can be
combined to achieve better fairness.

\textbf{Prior Work on Analytical Performance Modeling.} 
Prior works attempt to
quantify the impact of cache/memory contention through offline
profiling. Mars et al.~\cite{bubbleup} estimate an application's
sensitivity/propensity to receive/cause interference.  Other
previous works propose to estimate an application's
sensitivity to cache
capacity~\cite{cache-pirate,perf-variation-cache-sharing} and
memory bandwidth~\cite{bandwidth-bandit} through profiling.  Yang
et al.~\cite{bubbleflux} attempt to estimate applications'
sensitivity to interference online. However, this work assumes
that latency-critical applications run alone at times, when they
can be profiled (which could degrade system throughput).
These works assume the ability to profile
(1)~entire applications offline; or (2)~specific execution scenarios, such as an
application executing alone.  In contrast, MISE can estimate the slowdown of any
application \emph{online}, in the general scenario of multiple
applications running together.

Several previous
works~\cite{accurate-cpi-comp,mech-perf-model,superscalar-proc-model,pie}
propose analytical models to estimate processor performance,
as an alternative to time consuming simulations. The goal of our
\miiseabb model, in contrast, is to estimate slowdowns at runtime, in
order to enable mechanisms to provide QoS and high fairness. Its
use in simulation is possible, but is left to future work.

\section{Significance}
To
our knowledge, our HPCA 2013 paper~\cite{mise} is the first to build a \emph{simple yet
accurate} hardware\red{-based} model to estimate application
slowdowns due to main memory interference online \emph{with the
goal of providing predictable performance}. Previous
works~\cite{stfm, fst, cycle-accounting-taco,eiman-isca11}
propose mechanisms to estimate application slowdowns. However,
these mechanisms are not accurate enough (\red{as we demonstrate
in Section~\ref{sec:model-evaluation}}) since they were not
designed with the goal of providing predictable performance.
Rather, the slowdown estimates were used to make
prioritization/throttling decisions to improve overall fairness.


This work is also the first to design
a hardware\red{-based} mechanism to i) provide soft guarantees on
slowdown for applications and ii) detect when a prescribed
slowdown bound is not being met, by leveraging slowdown estimates
from the \miiseabb model, while also improving overall system
performance. Previous work~\cite{qos-sigmetrics}, in the absence
of a model to accurately estimate application slowdowns, always
prioritizes the application that needs guaranteed performance,
degrading the performance of other \red{co-running} applications.
Furthermore, previous work also does not have the provision to
detect whether or not the prescribed slowdown bounds are being met
(as we describe in Section~\ref{sec:leveraging}).

\subsection{Retrospective and Works Building on\\ Our HPCA 2013 Paper}
\label{sec:retrospective}

\textbf{Adoption of the Principles of the \miiseabb Model.}
The principles employed in the \miiseabb model have been adopted
towards slowdown estimation in several works that followed.  The
application slowdown model (ASM)~\cite{asm}, a follow\red{-}on
work, builds on top of \miiseabb's memory slowdown estimation model
and extended it to take into account shared cache interference. In
doing so, ASM also addressed one of the major caveats of the
\miiseabb model, the estimation of slowdown for
non-memory-intensive applications. While \miiseabb has a mechanism
to address the slowdown of non-memory-intensive applications, this
mechanism relies on the estimation of the memory\red{-}bound
fraction of an application. Estimating the fraction of an
application's execution that is memory bound, with high fidelity,
is challenging. ASM addresses this challenge by applying the
observation on request service rate as a proxy for performance at
the input to the shared caches. This seamlessly enables slowdown
estimation for applications with different memory and cache
intensities/sensitivities. \red{The ASM work shows that it can
accurately estimate slowdowns with only 9.9\% error across 100
workloads. We refer the reader to \cite{asm} for details.} 

A later work by Xiong et al.~\cite{sem} proposes a slowdown
estimation model that adopts the principle of giving an
application highest priority in order to estimate its alone run
behavior. This work directly measures alone\red{-}IPC during such
high priority periods, rather than estimating alone request
service rate and employs this alone\red{-}IPC estimate towards
determining slowdown. 



\textbf{Applications of the \miiseabb Model.}
\red{The MISE model has been applied towards slowdown estimation
in multiple contexts. Zhou and Wentzlaff~\cite{mitts} employ
the MISE model in the context of throttling memory traffic at the
source, based on inter-arrival times between requests.
Specifically, they employ a set of bins, each corresponding to
a range of inter-arrival times, and allocate a certain number of
credits to each bin, depending on an application's request
inter-arrival times. In order to determine the optimal credit
allocation in different bins corresponding to different arrival
times, they employ a genetic algorithm. This credit allocation
determines the eventual number of requests that can be served
corresponding to different inter-arrival times, for an
application, and hence, shapes the memory traffic of the
application. Slowdown estimates from the MISE model are leveraged
to determine the optimal bins/credits configuration, to
effectively shape memory traffic. 
Camouflage~\cite{camouflage}
 employs the MISE model for the purposes of
traffic shaping, but in the context of providing security.
Camouflage shapes memory traffic into
a predetermined distribution, in order to prevent attackers from
probing the memory bus to infer the program's memory access and
response patterns. Slowdown estimates from the MISE model are used
to determine the optimal bins/credits configuration.}

\indent\textbf{Employing Slowdown-Proportional Resource
Allocation.} The general principle of allocating resources
\red{proportionally}, to the estimated slowdown at that resource
is a key principle employed in the MISE-QoS and MISE-Fair schemes.
\ch{Two prior works~\cite{afp, xiang.ics17}} apply a similar principle in the
context of addressing interference at the on-chip network. Towards
mitigating on-chip network contention, they build a scheme that
allocates channel bandwidth proportional to the aggregate rate of
flow of traffic from each thread.      

These \red{works~\cite{asm,sem,mitts,camouflage,afp}} are clear
instances of the applicability of the MISE model itself and its
principles in various contexts. \red{The works that build on our
original MISE paper~\cite{mise}} are strongly indicative of the
potential impact this work could have in the long term, as we
describe in the next section.

\subsection{\red{Long-Term} Impact}

\textbf{Predictable Performance in
Current and Future Systems.} Building predictable systems is
a grand research challenge\ch{~\cite{cra, mutlu.superfri14, mutlu.imw13}}. 
Predictable performance is
a key requirement in current and future systems where \red{1)}
multiple applications are consolidated onto the same machine,
sharing resources \emph{and} \red{2)} some applications need
a certain guaranteed performance. Data centers, virtualized
systems, interactive mobile systems and real- time systems are all
examples of scenarios where predictable performance is desirable
or necessary. We expect the need for predictable performance to
increase in the future as more systems will likely move towards
consolidation as a means to effectively utilize resources. Given
this trend, accurately quantifying the effect of shared resource
interference on performance is an important enabler towards
providing predictable performance. Therefore, we believe that
slowdown estimates from the \miiseabb model and the
hardware/software techniques that can be built on top of our model
are important steps towards providing predictable performance.

\textbf{Request Service Rate a Proxy for Performance.} One
of the key ideas behind \miiseabb is to use memory request service
rate as a proxy for performance for memory-bound applications. We
hypothesize that the performance of an application that is
bottlenecked at a certain resource is likely correlated with the
request service rate at that resource. Hence, the notion of using
request service rate as a proxy for performance can be used as
a primitive for performance prediction and applied more generally
to other shared resources such as shared caches, storage and
network. \red{ASM~\cite{asm}, described in
Section~\ref{sec:retrospective}, is one such work that takes
advantage of this key idea of request service rate as a proxy for
performance, measured at the shared caches.}

\textbf{Accurate and Efficient Estimation of Alone
Performance.} Another key idea behind \miiseabb is to periodically
give each application the highest priority in order to estimate
\arsr. In doing so, the highest priority application receives
minimal interference when its slowdown is being estimated, while
also not disrupting other applications' execution. This leads to
better accuracy than previous work~\cite{stfm, fst,
cycle-accounting-taco,eiman-isca11} that estimates an
application's slowdown \emph{while} it is receiving interference
from other applications.  We believe that the principle of
estimating slowdown while using techniques such as prioritization
to minimize interference can be applied at other shared resources
such as I/O, \red{storage} and network as well.

%
\textbf{Enabling Better Resource Management.} The ability
to accurately estimate slowdown in the presence of shared resource
interference can enable a range of resource management techniques
to provide QoS in both hardware and software. Slowdown estimates
can be leveraged in the hardware for resource management (as we
demonstrate with memory bandwidth).  Slowdown estimates can also
be communicated to the software, enabling more effective and
informed admission control and migration mechanisms across
a cluster of machines. Therefore, we believe \miiseabb's slowdown
estimates can enable substantial future research on resource
allocation policies.

\textbf{Simplicity of the Technique.} The \miiseabb model
requires only simple hardware changes to the memory
\red{controller and scheduling} logic, while providing high
accuracy. By virtue of the memory bandwidth partitioning scheme we
employ, the memory scheduler only needs to give one application
the highest priority at any point in time, while treating other
applications' requests similarly. On the other hand, previously
proposed memory scheduling policies such as ATLAS,
TCM~\cite{atlas, tcm} employ ranking policies where an ordered
ranking is enforced across all applications' requests. Hence,
\miiseabb requires simpler comparator logic compared to previous
proposals and can be more easily incorporated into today's memory
controllers than previous proposals.

\textbf{\ch{Applicability to Other Memory Technologies.}}
\ch{In our HPCA 2013 paper~\cite{mise}, we described MISE within the context
of a system using DRAM as main memory, for which the reader can find
detailed background information in our prior works~\cite{atlas, tcm, salp, 
lee.hpca13, lee.hpca15, hassan.hpca16, liu.isca12, liu.isca13, chang.hpca14, 
chang.hpca16, seshadri.micro13, chang.sigmetrics17, lee.pact15, seshadri.micro17,
chang.sigmetrics16, hassan.hpca17, kim.cal15, lee.sigmetrics17, lee.taco16,
kim.isca14, patel.isca17, kim.hpca18}.
We believe the principles of MISE are easily applicable to other memory
technologies, e.g., phase-change memory~\cite{lee.isca09, qureshi.isca09, 
wong.procieee10, lee.ieeemicro10, zhou.isca09, lee.cacm10, yoon.taco14},
STT-MRAM~\cite{naeimi.itj13, kultursay.ispass13, meza.weed13},
and hybrid memory systems~\cite{qureshi.isca09, yoon.iccd12, phadke2011, rajeev2012, meza2012, li.cluster17,
meza.weed13, yu.micro17, ramos.ics11, zhang.pact09, agarwal.asplos17, 
dulloor.eurosys16, pena.cluster14, bock.iccd16, gai.hpcc16, liu.iccd16}.
We leave a detailed exploration of these to future works.}

\section{Conclusion}
\red{%
Application slowdowns induced by memory interference are
a significant deterrent to high and predictable performance.
Towards tackling such application slowdowns, 
our HPCA 2013 paper~\cite{mise}
(1)~builds a simple Memory Interference-induced Slowdown Estimation (MISE)
model to accurately estimate application slowdowns, and
(2)~demonstrates two use cases that leverage our MISE model to achieve
predictable performance and fairness. Since our original HPCA 2013
paper~\cite{mise} on the MISE model and its applications, several
works have adopted and employed the MISE model and its principles
in different contexts. We conclude that the MISE model and the
principles behind it can fuel and inspire many more such works on
high performance, predictable, and fair memory systems.}


\section*{Acknowledgments}
\red{
We thank Saugata Ghose for his dedicated effort in the preparation
of this article.
We thank the reviewers for their valuable feedback and
suggestions. We acknowledge members of the SAFARI group for their
feedback and for the stimulating research environment they
provide. Many thanks to Brian Prasky from IBM and Arup Chakraborty
from Freescale for their helpful comments. We
acknowledge the support of our industrial sponsors, including AMD,
HP Labs, IBM, Intel, Oracle, Qualcomm and Samsung. This research
was also partially supported by the NSF (grant 0953246), SRC,
and Intel URO Memory Hierarchy Program.
}

\bibliographystyle{IEEEtranS}
  \bibliography{references}
\end{document}